# Microscopic Characterization of the $L1_0$-FePt Nanoparticles Synthesized by the SiO$_2$-Nanoreactor Method

Yoshinori Tamada, Yasumasa Morimoto, Shinpei Yamamoto, Naoaki Hayashi[1], Mikio Takano, Saburou Nasu and Teruo Ono

*Institute for Chemical Research, Kyoto University, Uji 611-0011, Japan*
[1]*Department of Graduate School of Human and Environmental Studies, Kyoto University, Kyoto 606-8501, Japan*

We investigated magnetic properties of the $L1_0$-FePt nanoparticles synthesized by the "SiO$_2$-nanoreactor" method by means of Mössbauer spectroscopy from the microscopic point of view. Almost all of the nanoparticles were revealed to have nearly the same Mössbauer hyperfine parameters as those of the bulk $L1_0$-FePt alloy, indicating that they have well-defined $L1_0$ structure equivalent to the bulk state in spite of their small size of 6.5 nm.

KEYWORDS: FePt, nanoparticles, Mössbauer spectroscopy, nanocrystalline metal particles, fct-FePt, $L1_0$ phase, ferromagnetism, chemical synthesis

FePt alloy with the face-centered tetragonal $L1_0$ structure possesses high magnetic anisotropy energy ($K_u$, *ca.* $6 \times 10^6$ J/m$^3$), which is about an order of magnitude larger than that of the currently used CoCr-based alloys.[1,2] This strong magnetic anisotropy can suppress the superparamagnetic fluctuation of magnetization at room temperature down to a particle size of about 3 nm. Thus, FePt nanoparticles are expected as one of the most promising candidates for the future recording media with ultra-high densities beyond 1 Tbit/inch$^2$.[2,3] FePt nanoparticles synthesized by chemical solution-based methods attract much attention from the viewpoint of practical use.[4] This is because they have well-defined morphology and easiness to handle for the fabrication of desirable arrays on a substrate through being dispersible in solvents. However, the chemical solution-based methods can produce only disordered fcc or partially ordered $L1_0$-FePt nanoparticles,[5,6] and thus a post thermal annealing is necessary to transform them into the desired, well-ordered $L1_0$ structure. This post annealing results in coalescence and coarsening of the nanoparticles, leading to difficulties in fabricating desirable arrays on a substrate.

Recently we have succeeded in solving these problems by developing a new synthetic strategy named "SiO$_2$-nanoreactor" method.[7,8] Powder X-ray diffraction (XRD) and SQUID magnetometry studies revealed that the $L1_0$-FePt nanoparticles synthesized by this method possess well-crystallized $L1_0$ structure and a room temperature coercivity of 18.5 kOe, in spite of their small core size of only 6.5 nm in diameter. It was also demonstrated that they are dispersible in various solvents and the orientation of the magnetic easy axis of the solvent-dispersed nanoparticles can be controlled by an external magnetic field. Although the results of the XRD and the magnetization measurements indicated that the $L1_0$-FePt nanoparticles synthesized by SiO$_2$-nanoreactor method have good crystalline and magnetic properties, these techniques provide only macroscopic information. Characterization from the microscopic point of view is of great importance especially for nanoparticles which have a large number of surface atoms. Magnetic nanoparticles often exhibit different magnetic properties from bulk ones because of their special magnetic properties arising from structural imperfections near the surface.[9,10]

Here, we report the microscopic characterization of the $L1_0$-FePt nanoparticles synthesized by the "SiO$_2$-nanoreactor" method by means of the $^{57}$Fe Mössbauer spectroscopy. The Mössbauer spectroscopy is one of the useful tools for the investigation of the local electronic state of the Fe atom. It can reveal the existences of very small particles or surface oxidation layers which cannot be detected by the XRD. It was found that the $L1_0$-FePt nanoparticles synthesized by the "SiO$_2$-nanoreactor" method have well-defined $L1_0$ structure comparable to the bulk $L1_0$-FePt alloy.

$L1_0$-FePt nanoparticles were prepared according to the reported method.[7] The SiO$_2$-coated fcc-FePt nanoparticles were annealed at 900 ºC for 1 hr in flowing H$_2$(5%)/Ar(95%) gas to convert them to the $L1_0$ structure. Elemental composition of the FePt nanoparticles was determined to be Fe$_{52}$Pt$_{48}$ by using an atomic absorption spectrometer (Shimadzu, AA-6300). Figure 1 shows an XRD profile of the SiO$_2$-coated FePt nanoparticles after the annealing collected by using Cu $K_\alpha$ radiation ($\lambda$ = 0.154 nm) (Rigaku, RINT2500). The XRD profile shows the superlattice reflections (001), (110) and (002), which verify the formation of the ordered $L1_0$ structure. The degree of Fe/Pt ordering was estimated to be 0.85 from the relative intensity of the (110) and (111) peaks, which is similar to that of the previous works.[8] The mean crystallite size of the $L1_0$-FePt nanoparticles was estimated to be about 6.5 nm from half-maximum full width of the (111) peak by using the Scherrer formula. Figure 2 shows a transmission electron microscopic (TEM) image of the nanoparticles after the annealing (JEOL, JEM-1010D). This figure clearly shows that the nanoparticles were well isolated due to the $SiO_2$ coating, and the average core size was determined to be 6.7 nm.

Figure 3 shows the hysteresis loop of the $SiO_2$-coated $L1_0$-FePt nanoparticles measured at 5 K by using a Physical Property Measurement System (Quantum Design, PPMS) with an ACMS accessory. Here, $M_s$ represents the overall sample magnetization at 90 kOe. Because the amount of $SiO_2$ could not be determined precisely, it was impossible to make the magnetization specific with respect to the amount of the cores. Coercivity at 5 K reaches as large as 32 kOe, confirming Fe and Pt atoms forms well-ordered $L1_0$ structure. However, the hysteresis loop shows a kink at low magnetic field. This indicates the sample includes magnetically soft phase in addition to the $L1_0$ FePt phase. The amount of the soft magnetic phase was estimated to be 5% of all the nanoparticles from the decrease of magnetization at the kink.

The $^{57}$Fe Mössbauer measurements were performed in transmission geometry using a radioactive source of $^{57}$Co in Rh matrix with a constant acceleration technique. The velocity scale of the spectrum was relative to α-Fe at room temperature. The Mössbauer spectrum of the $L1_0$-FePt nanoparticles coated by $SiO_2$ at 5 K is shown in Fig. 4. The sharp sextet peaks of the spectrum strongly indicate the formation of the well-defined $L1_0$ phase with less fluctuation in composition as well as in crystalline structure. There is no signature of oxides within the experimental uncertainty. The spectrum can be well fitted with three subspectra: bulk-$L1_0$-FePt (site 1), surface-$L1_0$-FePt (site 2) and fcc-FePt (site 3). The values of isomer shift (IS), hyperfine field (HF), quadrupole splitting (QS), line width (WIDTH), and area fraction (AREA) of each subspectrum determined by the least square fit are listed in Table I. For the fitting, the line widths of the site 1 and the site 2 were fixed to the line width (0.31 mm/s) of the reference α–Fe at room temperature. The values of electric quadrupole interaction QS of the site 1 and the site 2 are in good agreement with that of the bulk $L1_0$-FePt reported by Goto *et al*.[11,12] The hyperfine field of the site 1 is about 10% larger than the reported value of the bulk $L1_0$-FePt at room temperature, which is quite reasonable by considering the difference in measurement temperatures. The fitting revealed that about 95% of the sample consists of the $L1_0$-FePt (site 1 and site 2). It is estimated from the area ratio of the site 1 to the site 2 that about 20 % of Fe atoms in the $L1_0$-FePt particle locates at the surface, which is reasonable for the nanoparticle with a diameter of 6.5 nm.

The other minor subspectrum (site 3) is assigned to the fcc-FePt, since the hyperfine parameters of the site 3 are in good agreement with the report values of the bulk fcc-FePt by Goto *et al*.[11,12] The Mössbauer spectrum can be fitted with higher accuracy by assuming the presence of the fcc-phase although its amount is very small (*ca*. 5 %). The fcc-FePt is known to show ferromagnetic behaviors with small coercivity at low temperature. Existence of a small amount of the magnetically isolated fcc-FePt nanoparticle can explain the kink of the hysteresis loop at low magnetic fields in Fig. 3. Recent report revealed that the fcc-FePt nanoparticles with sizes smaller than 4 nm cannot form the $L1_0$ structure upon annealing.[13] Since our synthetic method for the fcc-FePt nanoparticles is based on the seed-mediated growth mechanism and no attempt to suppress further nucleation was performed, thus-obtained fcc-FePt nanoparticles could contain such small-sized ones.

In conclusion, Mössbauer spectroscopic study revealed that the $L1_0$-FePt nanoparticles synthesized by the "SiO$_2$-nanoreactor" method have well-defined $L1_0$ structure comparable to the bulk state not only at the core but also near the surface. Only minor part (*ca*. 5%) of the sample was suggested to have the fcc structure, which could be attributed to the fcc-nanoparticles with such small sizes that cannot form the $L1_0$ structure. These too small fcc-nanoparticles might be produced during the course of synthesizing the fcc-FePt nanoparticles, and could thus be removed by adopting more elegant synthetic methods. Together with the results of the previous works,[7,8] the $L1_0$-FePt nanoparticles synthesized by this method are proved to have not only macroscopic but also microscopic magnetic properties equivalent to the bulk state in spite of their small size. With the advantage of being dispersible in various solvents, they are a promising material for the realization of future ultra-high density recording media.

**Figure Captions**

Figure 1
XRD profile of the $SiO_2$-coated FePt nanoparticles after the annealing.

Figure 2
TEM image of the $SiO_2$-coated FePt nanoparticles after the annealing.

Figure 3
Hysteresis loop at 5 K of the $SiO_2$-coated FePt nanoparticles after the annealing.

Figure 4
Mössbauer spectrum at 5 K of the $SiO_2$-coated FePt nanoparticles after annealing.

Table I

Mössbauer hyperfine values at 5 K of the SiO$_2$-coated FePt nanoparticles after the annealing. The values of isomer shift (IS) are relative to α-Fe at room temperature, not corrected for the second-order Doppler shift.

|  | IS (mm/s) | HF (T) | QS (mm/s) | WIDTH (mm/s) | AREA (%) |
|---|---|---|---|---|---|
| site 1 | 0.43 | 30.4 | 0.31 | 0.31 | 77.4 |
| site 2 | 0.43 | 28.5 | 0.31 | 0.31 | 17.3 |
| site 3 | 0.46 | 32.7 | 0.04 | 0.35 | 5.3 |

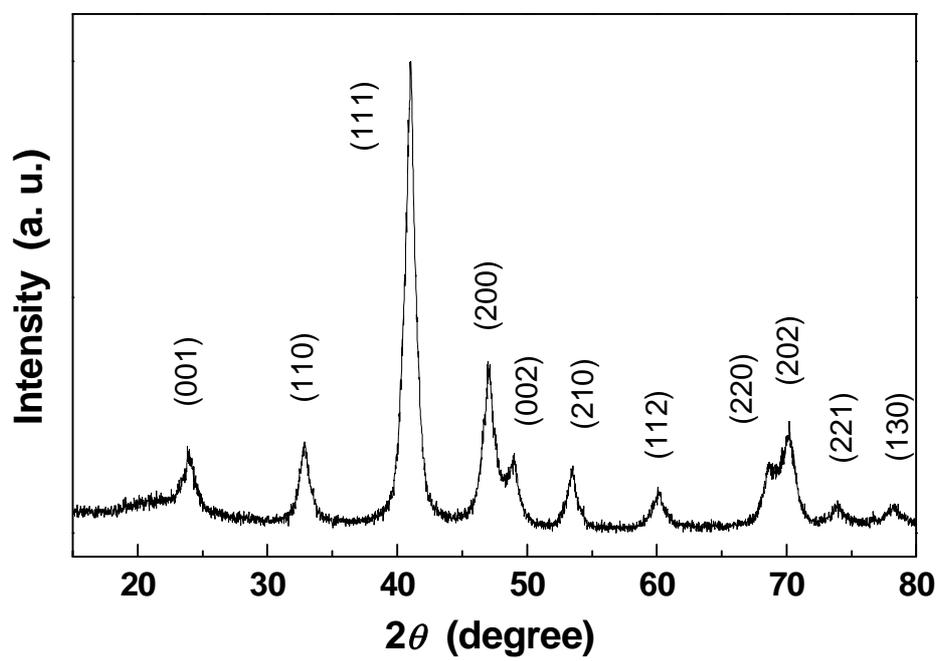

Figure 1
Y. Tamada

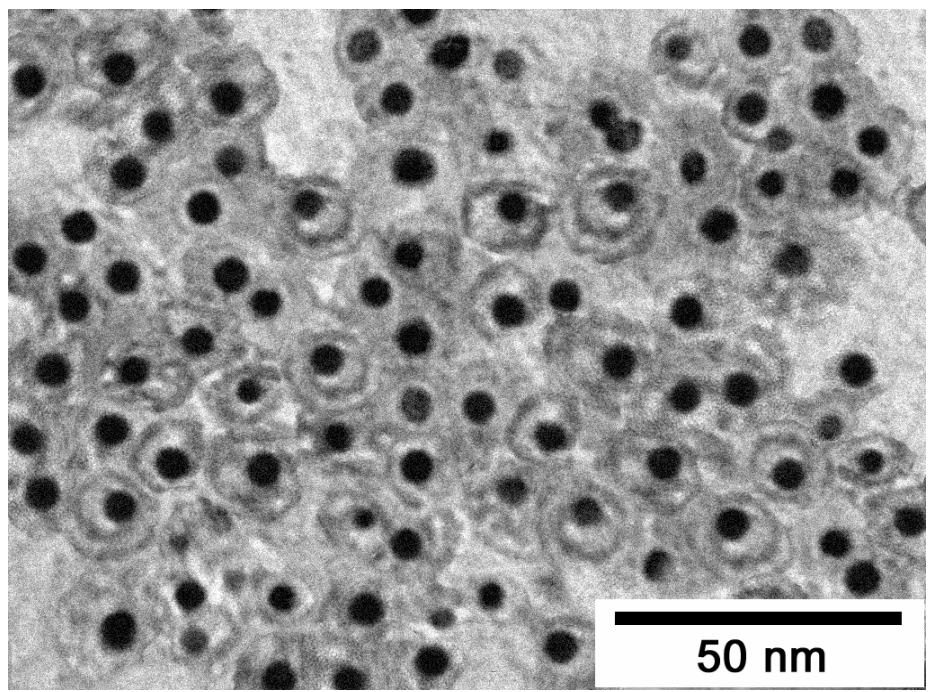

Figure 2
Y. Tamada

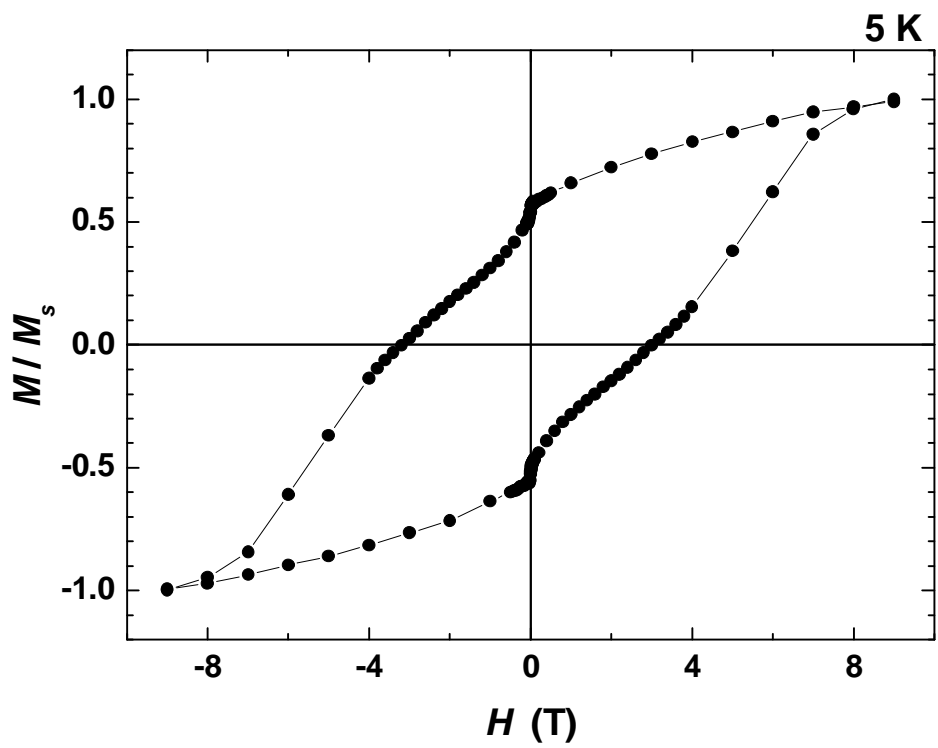

Figure 3
Y. Tamada

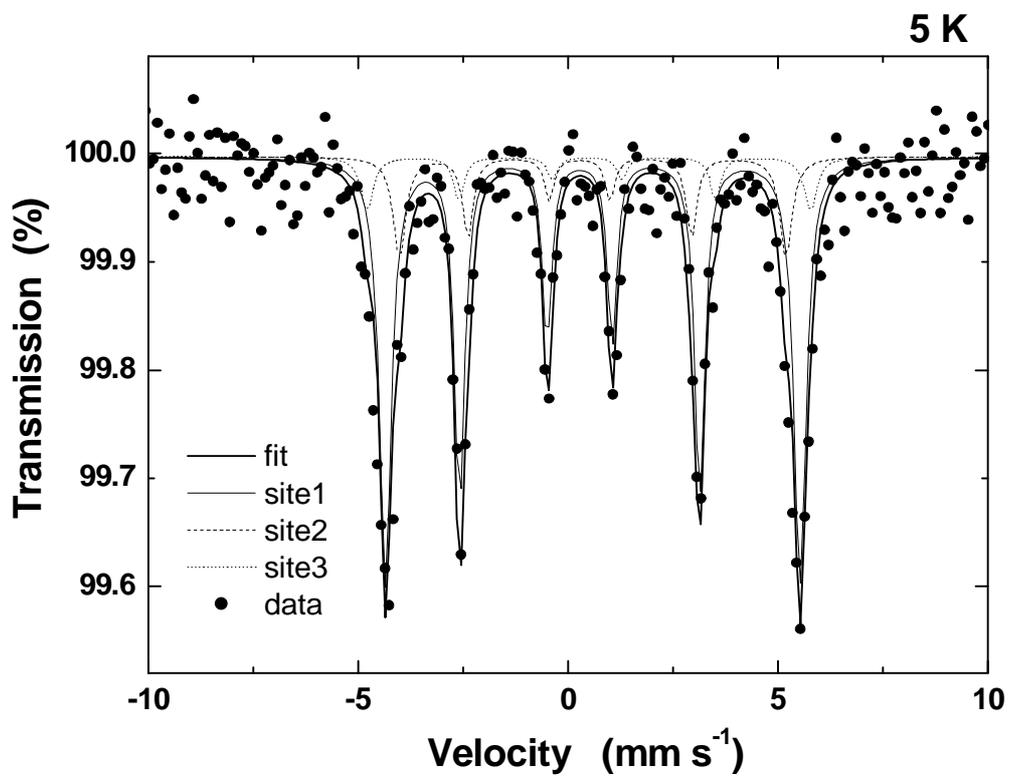

Figure 4
Y. Tamada